\documentstyle[epsf,proceedings]{crckapb}
\begin{opening}
\title{HYDRODYNAMICAL INSTABILITY OF EXTRAGALACTIC STRATIFIED JETS}
\author{M. HANASZ}
\institute{Institute of Astronomy, Nicolaus Copernicus University,
Toru\'n, Poland}
\author{H. SOL}
\institute{Observatoire de Paris, UPR 176 du CNRS, Meudon, France}
\end{opening}

\begin{document}

\section{Introduction}

A rather new view of extragalactic radio sources seems to come out from
recent observational data. For instance, in the case of 3C273, comparison
of optical and radio maps suggested the presence of an inner optical jet
embedded in a slow moving radio cocoon (Bahcall et al, 1995). Furthermore
new ways of combining radio data at different frequencies provided original
information on the sources. In particular sheaths of emission with
specific spectrum and polarization have been found around the jets in
Cygnus A, 3C449, two wide-angle-tail and one CSS sources, suggesting
that such kind of sheaths or envelopes are commonly present around
extragalactic jets (Katz-Stone, Rudnick, 1994; Rudnick, Katz-Stone,
1996).

The question of the origin of envelopes around extragalactic jets is
still open. They can just emanate from the inner jets as the result
of particle diffusion or correspond to some entrainment of the external
gas by the inner jets. Another possibility, suggested a few years ago in
the context of two-component jet models (Smith, Raine, 1985; Baker et
al, 1988; Sol et al, 1989; Achatz et al, 1990), is that they have been
directly ejected from the central engines. Indeed such models proposed
and analyzed the scenario where a fast beam (corresponding to the inner
core of the jet) is ejected from the vicinity of the central black hole
and a slower collimated outer wind (corresponding to the envelope)
is emitted by all parts of the accretion disk.

Anyway, such envelopes likely modify the interaction of the jets with the
ambient media in which they propagate, and a question which immediately
occurs is that of the Kelvin-Helmholtz stability properties of such
``core-envelope'' or ``stratified'' jets.

\section{Kelvin-Helmholtz instability of stratified jets}

We have studied the linear stage of the Kelvin-Helmholtz instability of
such composite stratified jets, in the plane parallel approximation
(Hanasz, Sol, 1996). Jets are assumed to be made of an inner core of
relativistic gas with a relativistic bulk velocity surrounded on both
sides by a sheet of non-relativistic gas with slower bulk velocity, and
embedded in the external ambient medium. There are therefore two types
of interfaces, the internal interfaces between the core and the sheet,
and the external ones between the sheet and the ambient medium. Transition
layers at all interfaces are described in the vortex-sheet
approximation. We perform a temporal stability analysis and perturb the
initial hydrodynamical equilibrium for the three components
(relativistic core and
non-relativistic sheet and external medium). The dispersion relation is
then obtained from the boundary conditions at the internal and external
interfaces, namely equality of pressure and equality of transverse gas
and interface velocities. It writes in compact form:
$$ {Z_s\over Z_c} {\cal E} = \coth \ ik_{c_\bot} \
 \  \hbox{for the symmetric solution} $$
and
$$ {Z_s\over Z_c} {\cal E} = {\rm th} \ ik_{c_\bot} \
 \  \hbox{for the antisymmetric solution.} $$
Here ${\cal E}$ is the term due to the effect of the envelope (or sheet),

$$ {\cal E} = {1+ {\cal R}_{se} e^{2ik_{s_\bot}(R-1)} \over
    1- {\cal R}_{se} e^{2ik_{s_\bot}(R-1)} } \ . $$
$R=R_s/R_c$ is the ratio of the sheet radius to the core radius,
$k_{c_\bot}$ and $k_{s_\bot}$
the transverse wavenumbers in core and sheet, $Z_c$, $Z_s$ and $Z_e$
the complex normal acoustic impedances of core, sheet and external medium.
${\cal R}_{se}$ is the reflection coefficient at the external interface
between sheet and ambient medium, defined as a function of the
acoustic impedances by
$$ {\cal R}_{se} = {Z_e -Z_s\over Z_e +Z_s} \ . $$
The dispersion relation reduces to the case of a single jet in different
limits, (i) when there is no reflection at the external interface
(${\cal R}_{se} =0$, ${\cal E}=1$), (ii) when the sheet becomes very thick
$(R>>1)$ since the factor of ${\cal R}_{s_e}$ then vanishes as
$\exp [-2R {\rm Im} (k_{s_\bot})]$ and ${\cal E}$ tends to 1 again,
(iii) when internal and external interfaces coincide $(R=1)$ and only the
core and external gas components remain.

\section{Results and interpretation}

The dispersion relation has been solved numerically and various sets
of parameters investigated. Figure 1 illustrates the temporal growth
rate obtained in a few cases. The presence of the sheet generates an
oscillating pattern due to the interferences of acoustic waves reflected
at the boundaries of the sheet, namely at the internal and external interfaces.
\begin{figure}
\epsfxsize=\hsize \epsfbox{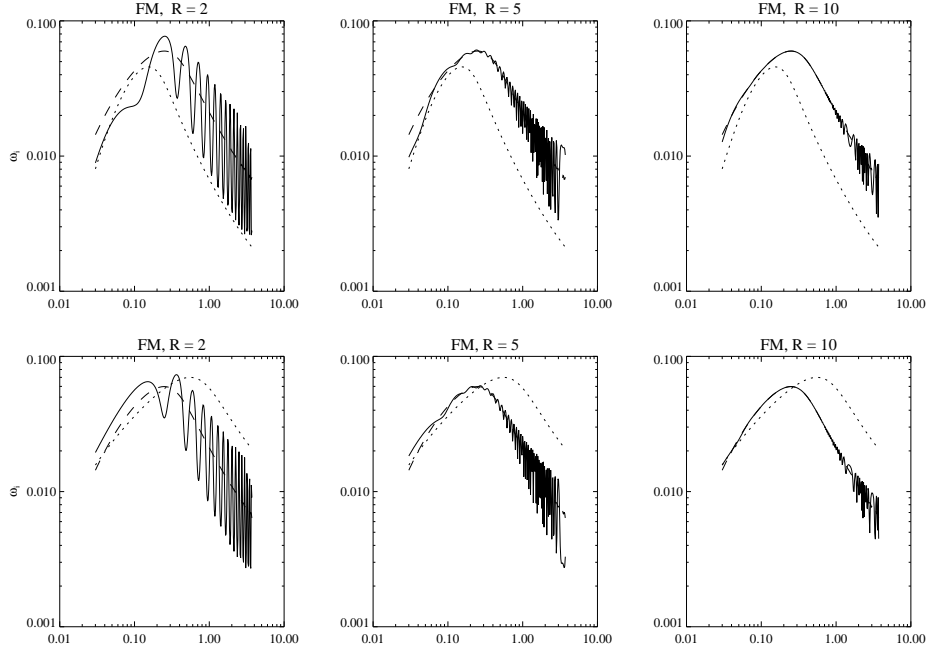}
\caption{Examples of temporal growth rate of the instability versus the
longitudinal wavenumber, for the fundamental mode of the antisymmetric
solution. The thickness of the sheet increases from first $(R=2)$ to
second $(R=5)$ and third $(R=10)$ columns. Upper case corresponds to
underdense core and sheet, with density ratios $\rho_c/\rho_s =0.01$
and $\rho_s/\rho_e =0.1$ and lower case to a dense sheet with
$\rho_c/\rho_s =0.01$ and $\rho_s/\rho_e =10$.
Here the relativistic bulk Lorentz factor of the core is 10 and the Mach
number of the sheet is zero. Dotted lines correspond to the solution
without sheet $(R=1)$ and dashed lines to the solutions without external
medium ($R\to \infty$, infinite sheet).}
\end{figure}

The oscillating pattern can be explained as follows. The acoustic waves
emitted at the internal interface are reflected at the external
interface and come back to the internal one. The reflection coefficient
for acoustic waves at external interface $|{\cal R}_{se}|$ is smaller
than 1, so there is no amplification of the wave amplitude there even if
sheet is supersonic (this statement is valid for frequencies and
wavenumbers typical for the core-sheet instability). The space between
internal and external interfaces (i.e. the sheet itself) can be
considered as the resonant cavity, which imposes the condition for resonance
of the type
$$ 2L = n \lambda_s $$
and antiresonance
$$ 2L = \left( n + {1\over 2} \right) \lambda_s $$
where $L$ is the distance between internal and external interfaces, and
$\lambda_s$ is the wavelength of the acoustic waves in the sheet,
both of them are measured along the direction of propagation.
The local maxima/minima of the value of growth rate correspond to the
fulfillment of resonant/antiresonant condition.
This is worth noticing that the resonant/antiresonant conditions are in
general modified by an additional phase shift coming from the reflection
at interfaces. The phase shift is dependent mostly on the density contrasts.
This is apparent in figure~1 where the upper and lower rows represent
two cases which differ only in the value of $\nu_s =\rho_s/\rho_e$
In the two cases the positions of maxima and minima are interchanged due to
the mentioned phase shift.
In all cases with thick enough sheet $(R>1)$ the solutions of the dispersion
relation rapidly converge to the case
of infinite sheet. So a sheet or envelope as thick as the inner core almost
imposes the growth rate. Parameters of the envelope rather than those
of the external medium appear to determine the Kelvin-Helmholtz instability,
and the inner beam is somewhat isolated from the ambient gas. Relatively
to the solution without envelope, the growth rate is increased for an
envelope light  compared to the external medium, and lowered for a dense
envelope. So the presence of an heavy envelope can help to stabilize
the jet configuration.
\begin{quote}

\end{quote}

\end{document}